# ProME: An Integrated Computational Platform for Material Properties at Extremes and Its Application in Multicomponent Alloy Design

*Xingyu Gao, William Yi Wang, Xin Chen, Xiaoyu Chong, Jiawei Xian, Fuyang Tian, Lifang Wang, Huajie Chen, Yu Liu, Houbing Huang, and HaiFeng Song\**

Xingyu Gao, Xin Chen, Jiawei Xian, Lifang Wang, Yu Liu, HaiFeng Song
National Key Laboratory of Computational Physics, Institute of Applied Physics and Computational Mathematics, Beijing 100088, China
E-mail: song_haifeng@iapcm.ac.cn

William Yi Wang
State Key Laboratory of Solidification Processing, Northwestern Polytechnical University, Xi'an 710072, China

Xiaoyu Chong
Faculty of Materials Science and Engineering, Kunming University of Science and Technology, Kunming 650093, China

Fuyang Tian
Institute for Applied Physics, University of Science and Technology Beijing, Beijing 100083, China

Huajie Chen
School of Mathematical Sciences, Beijing Normal University, Beijing 100875, China

Houbing Huang
School of Multidisciplinary Science, Beijing Institute of Technology, Beijing 100081, China

Keywords: extreme conditions, integrated computation, multicomponent alloys, wide temperature-pressure range, high strain rates

We have built an integrated computational platform for material properties at extreme conditions, ProME (Professional Materials at Extremes) v1.0, which enables integrated calculations for multicomponent alloys, covering high temperatures up to tens of thousands of Kelvin, high pressures up to millions of atmospheres, and high strain rates up to millions per second. A series of software packages have been developed and integrated into ProME v1.0, including ABC (AI-Based Crystal search) for crystal structure search under pressure, SAE (Similar Atomic Environment) for disordered configuration modeling, $MFP^2$ (Multiphase Fast Previewer by Mean-Field Potential) for multiphase thermodynamic properties, HTEM (High-throughput Toolkit for Elasticity Modeling) for thermo-elastic properties, TREX (TRansport at Extremes) for electrical and thermal conductivity, Hippos (High plastic phase model software) for phase-field simulation of microstructure evolution under high strain rates, and AutoCalphad for modeling and optimization of phase diagrams with variable compositions. ProME v1.0 has been applied to design the composition of the quaternary alloys Platinum-Iridium-Aluminum-Chromium (Pt-Ir-Al-Cr) for engine nozzles of aerospace attitude-orbit control, achieving high-temperature strength comparable to the currently used Pt-Ir alloys but with significantly reduced costs for raw materials. ProME offers crucial support for advancing both fundamental scientific understanding and industrial innovation in materials research and development.

## 1. Introduction

Understanding the physical properties of materials under extreme conditions, such as high temperatures, high pressures, and high strain rates, is essential for the design and optimization of crucial components in fields like aerospace and power industry. However, performing experiments to measure material properties under such conditions is quite challenging, due to high costs, long duration, measurement difficulties, and etc. Furthermore, simulations confined to a limited spatio-temporal scale often fail to capture the multiple factors that influence service performance and the evolution of these factors. Therefore, it is of great significance to construct an integrated computational platform capable of predicting material properties and performance under extreme conditions. The development of such a platform would help to accelerate the design of novel materials and enhance the performance and reliability of equipment.

Determining the phase structure is a fundamental step in studying material properties under extreme conditions, primarily involving crystal structure search and disordered configuration modeling. Established methods and software for crystal structure search include USPEX[1] based on genetic algorithms, CALYPSO[2] based on particle swarm optimization, and STEPMAX[3] based on basin-hopping algorithms. In recent years, machine learning force fields (MLFFs) have been increasingly used to accelerate *ab initio* structural relaxation, often the computational bottleneck in these searches, but achieving an optimal balance between efficiency and reliability remains a challenge. For modeling disordered configurations, such as those found in high-entropy alloys, one approach involves effective medium methods like the virtual crystal approximation (VCA)[4] and the coherent potential approximation (CPA).[5] Another employs supercell-based methods, including the cluster expansion (CE) method[6,7] and the special quasi-random structures (SQS) method.[8] Recently, Song *et al.* developed the supercell-based similar atomic environment (SAE) method,[9] which provides an effective alternative to SQS for generating quasi-random structures.[10]

Free energy plays a central role in the calculation of thermodynamic properties of materials. At low and moderate temperatures, the quasi-harmonic approximation (QHA)[11] is well-established for the prediction of thermodynamic properties of solids. As temperature rises, lattice anharmonic vibrational effects become more pronounced, and may have a large influence on the thermodynamic properties, such as solid-solid and solid-liquid phase boundaries. To account for anharmonic effects on the free energy, two categories of computational methods have been developed. One relies on analyzing the vibrational density of states, including methods for solids such as the phonon quasi-particle spectra (PQS) method,[12] the temperature-dependent effective potential (TDEP) method,[13] and the self-consistent *ab-initio* lattice dynamics (SCAILD) method,[14] as well as the two-phase thermodynamic (2TP) model for liquids.[15] The other is based on the formally exact thermodynamic integration technique, such as the unsampled thermodynamic integration using Langevin dynamics (UPTILD) method[16] and the harmonically mapped averaging (HMA) method.[17] However, applying these methods to complex systems like multicomponent alloys can be challenging due to the need for computationally expensive molecular dynamics simulations[12–13, 15–17] or numerous static calculations with large supercell configurations.[14]

Predicting material properties under high strain rates requires consideration of microstructural evolution, which is difficult to describe entirely at the microscale. To bridge this gap, the phase-field method simulates the evolution of material microstructures at the mesoscale by minimizing free energy and can be coupled with material mechanics models. It has been increasingly applied to calculate material properties involving mesoscale microstructure, deformation, and mechanical characteristics.[18,19] Shi *et al.* improved the traditional micro-elasticity theory by coupling microstructural evolution with plastic deformation,[20] describing near-equilibrium static high-pressure loading processes. Liu *et al.* coupled phase-field simulations with crystal plasticity finite element simulations to study microstructure and deformation features under various loading conditions.[21] However, this coupling between microstructure and deformation is considered "loose", deviating from actual conditions and impacting computational efficiency.

Phase diagrams are fundamental for predicting material properties and phase stability. Currently, several commercial software packages exist for phase diagram calculations, such as Thermo-Calc,[22] Pandat,[23] FACTSage,[24] and JMatPro.[25] These tools primarily perform calculations based on existing thermodynamic databases. However, developing thermodynamic databases for new material systems often relies heavily on empirical fitting and remains a significant challenge. To address this, Liu *et al.* developed ESPEI,[26] an automatic optimization program for thermodynamic model parameters, which has been successfully applied to multicomponent alloys.[27–29] Despite its advances, ESPEI has not yet achieved fully automated integration with first-principles calculation software and requires further extension to accurately describe high-temperature and high-pressure phase diagrams.

Regarding integrated computational platforms, the JARVIS platform integrates classical force field calculations, first-principles simulations, and machine learning algorithms to achieve multi-scale prediction of basic material properties through standardized data interfaces.[30] The MATERIALS PROJECT platform, relying on a high-throughput computational framework, integrates structure prediction, phase diagram generation, and thermodynamic databases, forming an important infrastructure for the Materials Genome Initiative.[31] The CHiMaD platform, by integrating phase-field simulations, elasticity calculations, and intermetallic compound databases, possesses multi-physics coupling analysis capabilities.[32] In recent years, under the support of the China Materials Genome Initiative, the ALKEMIE[33] and MatCloud[34] platforms have been developed to construct visual workflow of integrated computation using various simulation softwares and to perform analysis of the associated computational results, and the CNMGE 1.0 platform[35] has built an autonomous and controllable cross-domain efficient resource scheduling system to enable high-throughput materials computation simulation and data management. Nevertheless, existing platforms, in terms of both computational methods and data accumulation, cannot meet well the demands posed by extreme conditions.

To address the challenges in calculating material properties under extreme conditions (high temperatures, high pressures, and high strain rates) and constructing an integrated computational platform, we have developed a series of key techniques: an on-the-fly machine learning force-field accelerated crystal structure search algorithm, a high-temperature free energy model based on modified mean-field potential (MMFP), a phase-field model for grain evolution and mechanical response under high strain rates, and integrated computational modeling and automatic optimization techniques for the phase diagrams of multicomponent alloys. Based on these developments, we established the integrated computational platform for multicomponent alloys, ProME (Professional Materials at Extremes) v1.0. As shown in Figure 1, this platform is applicable to extreme conditions covering high temperatures (up to tens of thousands of Kelvin), high pressures (up to millions of atmospheres), and high strain rates (up to millions per second). Using ProME, we have achieved an efficient computational workflow of "material composition - phase structure - extreme properties". Within half a year, we completed the compositional design of the quaternary alloys Platinum-Iridium-Aluminum-Chromium (Pt-Ir-Al-Cr) for engine nozzles of aerospace attitude-orbit control, achieving high-temperature strength comparable to the currently used Pt-Ir alloys but with significantly reduced costs for raw materials. Such tasks might previously have required several years to complete.

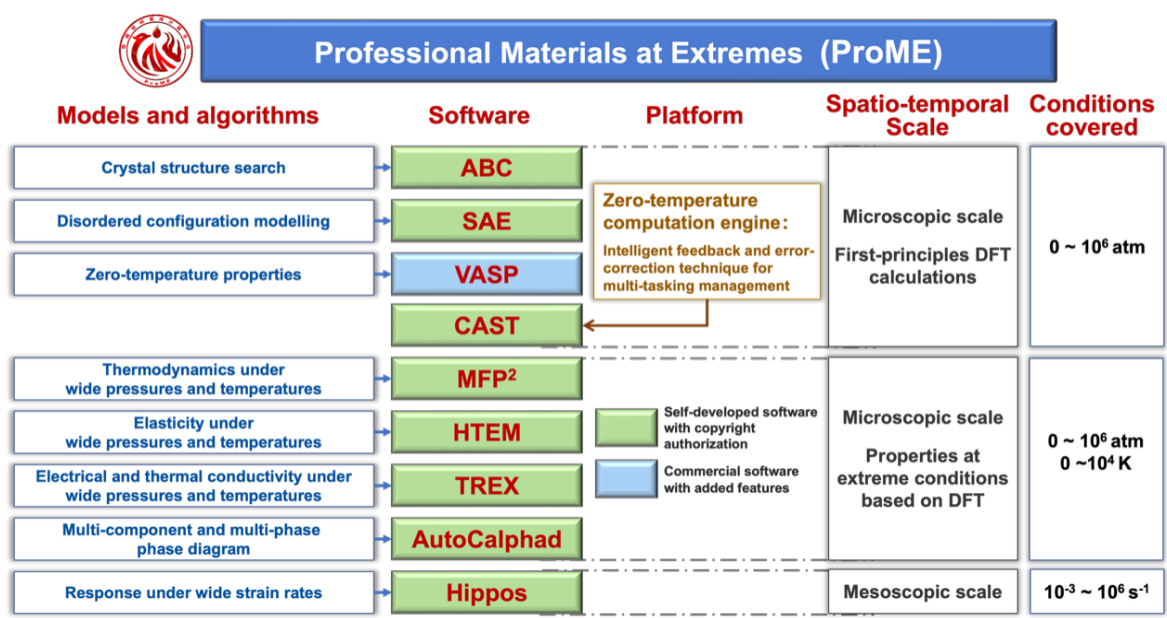

**Figure. 1** The architecture of the computational platform ProME.

## 2. High-Pressure Crystal Structure Search and Disordered Configuration Modeling

Investigating the properties of materials under extreme conditions first necessitates determining their stable phase structures at relevant pressures, typically achieved through crystal structure search,[36] where enhancing the search efficiency remains a key challenge. Crystal structure search typically requires numerous structural relaxations, often performed using computationally expensive *ab initio* methods. The total computational cost is thus proportional to the product of the number of search steps and the average cost per relaxation. Initial efforts focused on reducing the number of search steps, as implemented in software like

CALYPSO, USPEX, and TGMin, which utilize efficient global optimization algorithms and symmetry-based constraints.[1, 2, 37–39] Subsequently, MLFFs have been employed to reduce the computational cost per relaxation.[40] However, pre-training an accurate MLFF demands substantial computational resources. Furthermore, because structure searching inherently involves exploring vast, unknown regions of the configurational space, the extrapolation reliability of pre-trained MLFFs becomes uncertain. Although updating MLFFs on-the-fly during the search process was attempted by Hammer *et al.* to accelerate simulations, their force field formalism was relatively simple and less suited for the complexities of high-pressure crystal structure prediction.[41–43] Balancing the cost of force field development, the total structural relaxation cost, and the need for search reliability, we developed the ABC software, which utilizes on-the-fly MLFF acceleration tailored for these demanding conditions. Based on the crystal structures identified by ABC, we then employ the SAE software to model the corresponding chemically disordered configurations.

## 2.1. ABC for High-Pressure Crystal Structure Search

Figure 2(a) illustrates the architecture of the ABC software, centered around an on-the-fly machine learning force field (MLFF) update module. This module utilizes a modified version of the TensorMD force field,[44–46] which is dynamically generated and updated during the structure search to pre-relax candidate structures. (A detailed description of this algorithm is provided in the Supporting Information). To support these on-the-fly updates, ABC incorporates a comprehensive post-processing module. This module analyzes each computational task, recording detailed metadata, including structural descriptors, space groups, pressure, atomic forces, and computational cost, which informs subsequent MLFF refinements. Furthermore, recognizing that many materials adopt high-symmetry structures under pressure, ABC implements a symmetry-constrained configuration sampling method based on depth-first search and combinatorial algorithms[38] (details in the Supporting Information).

ABC employs a three-stage relaxation strategy designed to balance computational cost with result reliability: (1) Perform symmetry-constrained relaxation using the on-the-fly MLFF on initially generated random configurations. (2) For configurations that fail to converge in the first stage, perform unrestricted relaxation using the MLFF. (3) Utilize the MLFF-relaxed configurations as starting points for final refinement with first-principles calculations. This hierarchical approach leverages the speed of MLFFs and the efficiency gains from symmetry constraints to rapidly identify promising structures, while ensuring the accuracy and reliability of the final results through first-principles refinement. Figure 2(b) displays the search results for $Mg_1Al_3$ at 100 GPa, demonstrating consistency with previous studies.[47] Figure 2(c) quantifies the acceleration achieved by the on-the-fly MLFF: an initial MLFF is constructed after 10 steps and subsequently updated every 20 steps. Compared to relying solely on first-principles calculations, this approach identified the identical ground-state and key metastable structures while reducing the total CPU time by 58%.

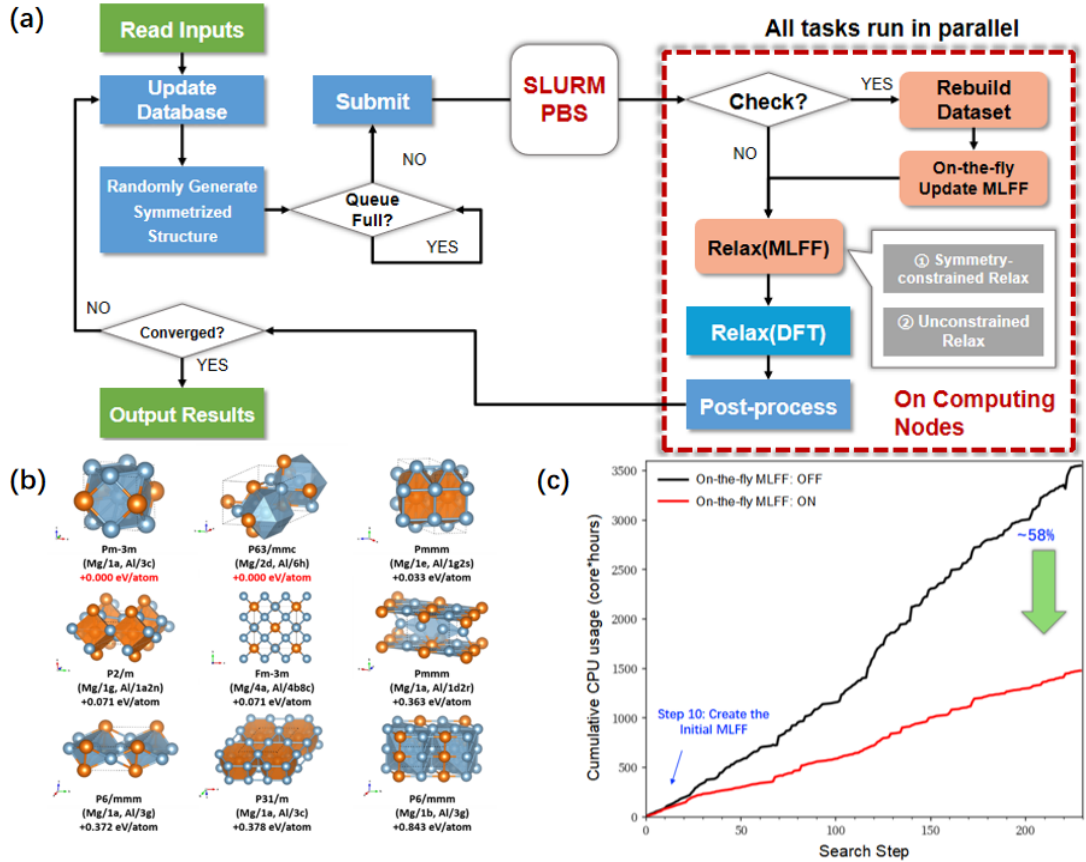

**Figure 2.** (a) Architecture of the ABC software. (b) Stable and metastable structures identified for the $Mg_1Al_3$ system at 100 GPa.[47] (c) Comparison of computational acceleration using the on-the-fly MLFF.

## 2.2. SAE for Disordered Configuration Modeling

Once the underlying crystal structure is determined, we employ the SAE method[9] to model disordered atomic arrangements, such as those in metal alloys.[8] The SAE method utilizes a similarity function to quantify the deviation between the atomic environment within a finite-sized supercell and that of an ideal random configuration. By extending this concept with short-range order parameters to construct a generalized similarity function, the method can uniformly describe both overall chemical disorder and local short-range order effects. SAE has been successfully applied to model atomic configurations in various systems, including multicomponent alloys and materials exhibiting short-range order, such as certain high-entropy alloys.[48–54] By analyzing the generation time for 500 binary body-centered cubic (BCC) disordered configurations, we examined the scaling of computational time with system size. As shown in Figure 3, the SAE method demonstrates linear scaling complexity. In contrast, the Special Quasi-random Structures (SQS) method, as implemented in the ATAT software,[8] involves global updates of the atomic environment (partly due to allowing variations in supercell vectors), leading to approximately quadratic scaling complexity. Although SAE performs local updates and does not explicitly optimize cell vectors, cross-validation, as illustrated in Table 1, indicates that the quality of disordered configurations generated by SAE is comparable to those produced by SQS.

**Table 1.** Cross-validation of modeling quality for SAE versus SQS: Following typical cluster expansion (CE) cutoff conventions[55-59] (e.g., two-body correlations often extend up to the 5th or 6th nearest neighbor, while three-body correlations typically extend up to the 2nd nearest neighbor), the quality of disordered configurations generated by 10,000 Monte Carlo sampling steps using SAE and SQS, respectively, is compared by examining the resulting two-body ($\Delta_{2,\text{SAE}}$, $\Delta_{2,\text{SQS}}$) and three-body ($\Delta_{3,\text{SAE}}$, $\Delta_{3,\text{SQS}}$) cluster correlation functions.

| Symmetry | Composition | Cutoff (Å) | SAE structure | | | | SQS structure | | | |
| --- | --- | --- | --- | --- | --- | --- | --- | --- | --- | --- |
| | | | $\Delta_{2,\text{SAE}}$ | $\Delta_{3,\text{SAE}}$ | $\Delta_{2,\text{SQS}}$ | $\Delta_{3,\text{SQS}}$ | $\Delta_{2,\text{SAE}}$ | $\Delta_{3,\text{SAE}}$ | $\Delta_{2,\text{SQS}}$ | $\Delta_{3,\text{SQS}}$ |
| BCC | AB | 0.9 | 0.000 | 0.000 | 0.000 | 0.000 | 0.000 | 0.000 | 0.000 | 0.000 |
| | | 1.05 | 0.006 | 0.020 | 0.011 | 0.000 | 0.006 | 0.019 | 0.011 | 0.021 |
| | ABC | 0.9 | 0.000 | 0.000 | 0.000 | 0.000 | 0.000 | 0.000 | 0.000 | 0.000 |
| | | 1.05 | 0.000 | 0.000 | 0.000 | 0.000 | 0.011 | 0.030 | 0.000 | 0.031 |
| FCC | AB | 0.75 | 0.000 | 0.000 | 0.000 | 0.000 | 0.000 | 0.000 | 0.000 | 0.000 |
| | | 1.05 | 0.000 | 0.007 | 0.000 | 0.007 | 0.000 | 0.009 | 0.000 | 0.002 |
| | ABC | 0.75 | 0.000 | 0.000 | 0.000 | 0.000 | 0.000 | 0.008 | 0.000 | 0.003 |
| | | 1.05 | 0.000 | 0.007 | 0.000 | 0.013 | 0.000 | 0.063 | 0.000 | 0.046 |
| HCP | AB | 1.05 | 0.012 | 0.021 | 0.025 | 0.000 | 0.012 | 0.021 | 0.025 | 0.000 |
| | | 1.5 | 0.012 | 0.042 | 0.037 | 0.000 | 0.013 | 0.019 | 0.031 | 0.005 |
| | ABC | 1.05 | 0.000 | 0.011 | 0.000 | 0.064 | 0.021 | 0.182 | 0.021 | 0.205 |
| | | 1.5 | 0.023 | 0.124 | 0.020 | 0.127 | 0.016 | 0.219 | 0.013 | 0.195 |

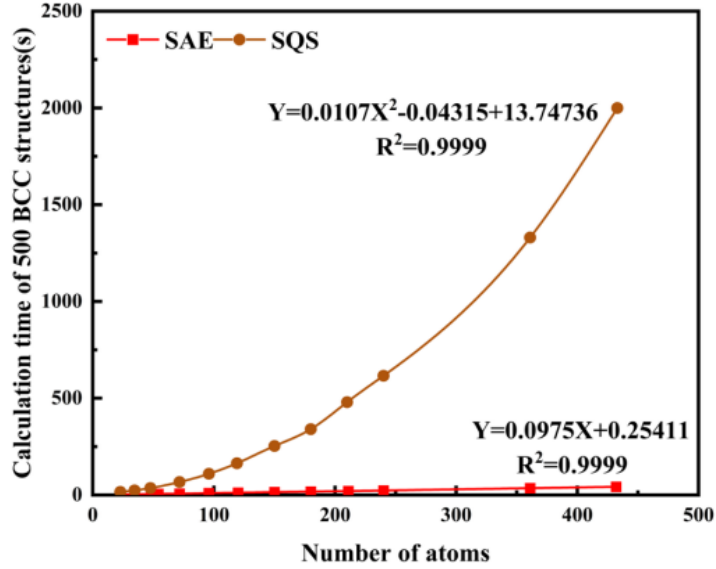

**Figure 3.** Comparison of modeling efficiency between SAE and SQS methods.

## 3. Material Properties under High-Temperature Conditions

Under high-temperature and high-pressure conditions, materials may undergo phase transitions (e.g., solid-solid, solid-liquid), leading to abrupt changes in their physical properties. Accurately capturing these phase transitions is critical for understanding material behavior under such extreme conditions. While the quasi-harmonic approximation (QHA)[11] works well for the calculation of thermodynamic properties of solids at low and moderate temperatures, it neglects the contribution of lattice anharmonic vibrations to the free energy, which may become important at high temperatures. To address this, we improved the mean-field potential (MFP) method to provide a new model for rapid free energy calculations. Based on this, we developed the MFP$^2$ (Multiphase Fast Previewer by Mean-Field Potential) software,[60] which properly describes both the low-temperature quantum effects and the high-temperature anharmonic effects, with a computational cost comparable to QHA. With the ability to determine phase boundaries from free energy calculations established, we also developed software for calculating other thermal properties, including the HTEM (High-throughput Toolkit for Elasticity Modeling) software[61–63] for thermos-elasticity and the TREX (TRansport at Extremes) software[64, 65] for electrical and thermal conductivity.

### 3.1. MFP$^2$ for Multiphase Thermodynamic Properties

The MFP method constructs a mean-field potential based on the cold-energy curve obtained from first-principles calculations. This potential is then used to rapidly determine the contribution of ionic vibrations to the free energy, with the high-temperature anharmonic effects partially accounted for.[66, 67] In Table 2, comparison is made between different models for the calculation of ion-vibrational free energy. We can see that the traditional MFP method does not consider the low-temperature quantum effects, limiting its accuracy in the low-temperature regime. A PG-type correction[67] was proposed to resolve this, but from low-temperature asymptotic analysis it was shown that there are still deficiencies in describing quantum effects. We proposed a modified MFP (MMFP) method[60] which successfully addressed this issue. In MMFP, a characteristic temperature $\theta_\lambda(V)$ is first derived from the generalized force constants in the MFP method (details in the Supporting Information), where $\lambda$ represents a parameter that controls the coupling between the Grüneisen parameter and the cold-energy curve in MFP. Then, the ion-vibrational free-energy difference between QHA and the high-temperature Debye model with Debye temperature $\theta_\lambda(V)$ is used to correct the MFP free energy. This leads to an expression for the ion-vibrational free energy which accurately describes the low-temperature quantum effects and partially captures the high-temperature anharmonic effects

$$F_{\text{ion}}^{\text{MMFP}}(V,T) = F_{\text{ion}}^{\text{MFP}}(V,T) + F_{\text{ion}}^{\text{QHA}}(V,T) - 3k_\text{B} T \ln\left[\frac{\theta_\lambda(V)}{T}\right], \quad (1)$$

where $F_{\text{ion}}^{\text{MFP}}(V,T)$, $F_{\text{ion}}^{\text{QHA}}(V,T)$ and $3k_\text{B} T \ln[\theta_\lambda(V)/T]$ are the ion-vibrational free energy from MFP, QHA, and the high-temperature Debye model, respectively.

**Table 2.** Comparison between different models for the calculation of ion-vibrational free energy.

| Model | Formula of ion-vibrational free energy | Low-temperature quantum effects | High-temperature anharmonic effects |
|---|---|---|---|
| Mean-field potential (MFP) | $F_{\text{ion}}^{\text{MFP}}[\theta_\lambda(V), T]$, $\theta_\lambda(V) = \frac{\hbar}{k_\text{B}}\sqrt{\frac{2k_\lambda(V)}{M}}$ | No quantum effects | Partial anharmonic effects |
| Quasi-harmonic approximation (QHA) | $F_{\text{ion}}^{\text{QHA}}[\theta_0(V), T]$, $\ln(k_\text{B}\theta_0) = \langle\ln(\hbar\omega)\rangle$ | Quantum effects | No anharmonic effects |
| PG-type correction[67] | $F_{\text{ion}}^{\text{MFP}}[\theta_\lambda(V), T] + F_{\text{ion}}^{\text{QHA}}[\theta_0(V), T]$ $- 3k_\text{B} T \ln\left[\frac{\theta_0(V)}{T}\right]$ | Quantum effects+$o(T)$ | Partial anharmonic effects[66, 67] |
| Modified mean-field potential (MMFP) | $F_{\text{ion}}^{\text{MFP}}[\theta_\lambda(V), T] + F_{\text{ion}}^{\text{QHA}}[\theta_0(V), T]$ $- 3k_\text{B} T \ln\left[\frac{\theta_\lambda(V)}{T}\right]$ | Quantum effects+ $o[\sqrt{k_B T}\exp(-\kappa(V)\frac{\theta_\lambda^2(V)}{T})]$ | Partial anharmonic effects[66, 67] |

Based on the MMFP method, we developed the MFP² software for calculating multiphase thermodynamic properties. The main workflow consists of three stages, as shown in Figure 4: (1) Calculations are performed using first-principles simulation software (e.g., VASP[68]) based on the density-functional theory (DFT)[69] to obtain basic data for subsequent calculations in the MFP$^2$ software, including cold energy, phonon spectra, and electronic density of states. (2) From these basic first-principles data, the ion-vibrational free energy is determined from the mean-field potential constructed in the MFP method and corrected using our proposed MMFP method, and subsequently the total Helmholtz free energy can be obtained. (3) Using the relationship between the Helmholtz free energy and other thermodynamic quantities, solid-solid and solid-liquid phase boundaries, as well as thermodynamic properties of different phases, can be derived from the calculated Helmholtz free energy.

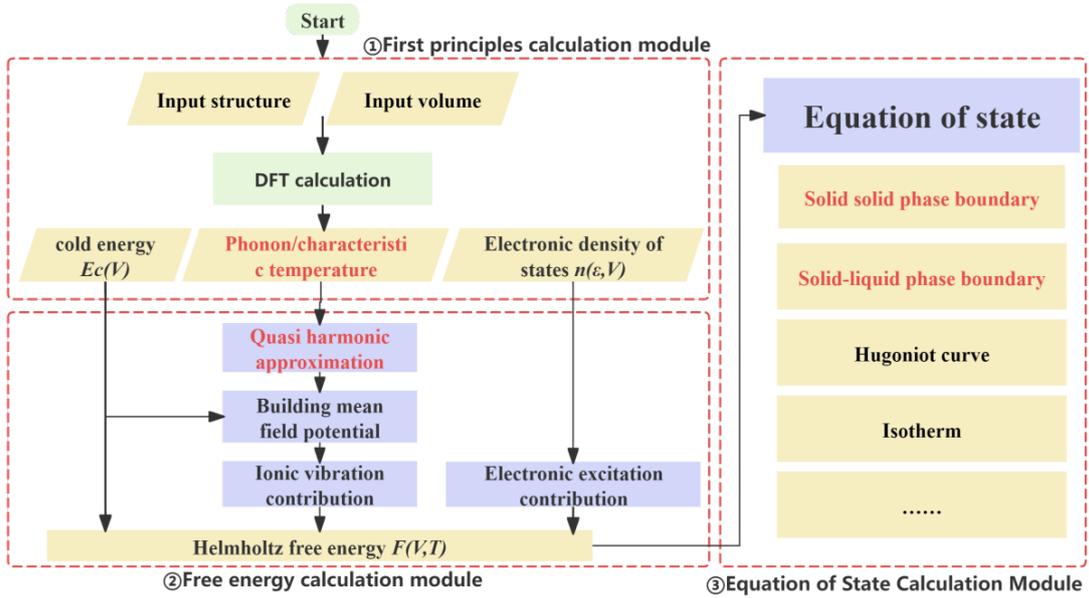

**Figure 4.** Workflow of the MFP$^2$ software.

We validated the accuracy of MFP$^2$ in calculating phase boundaries by taking the metallic Mg and a typical Mg-Al alloy AZ31B (with Mg and Al as major components) as prototypical examples. In the MFP$^2$ software, the high-pressure melting curve is obtained by combining the Lindemann criterion for melting[70] with the free energy from the MFP method. Figure 5 compares the phase boundaries of Mg and AZ31B calculated using MFP$^2$ with other first-principles calculations and experimental measurements.

Figure 5(a) presents the results for the phase boundary of Mg. The solid-solid hcp-bcc phase boundary from the MMFP method shows overall good agreement with experimental measurements.[71, 72] Furthermore, at relatively low temperatures, it is consistent with the QHA method, while at high temperatures when anharmonic effects become significant, it agrees well with the high-accuracy phonon quasi-particle spectra (PQS) method.[12] The melting curve from the MFP method generally agrees well with high-accuracy solid-liquid coexistence simulations.[73] At relatively low pressures, the melting curve from MFP agrees well with experiments.[74–76] At pressures above 50 GPa, experimental results for melting exhibit larger discrepancies,[72, 76, 77] possibly due to different criteria adopted for the onset of melting, kinetic effects, non-hydrostaticity, strength effects, or other effects.[77, 78] In particular, the results for melting from MFP are consistent with some of the experimental studies.[76, 77]

Figure 5(b) presents the results for the phase boundary of AZ31B. The hcp-bcc phase boundary from MMFP is consistent with QHA results at relatively low temperatures, and agrees well with the high-accuracy PQS method at high temperatures when anharmonic effects become prominent. The melting curve from the MMFP method shows excellent agreement with high-accuracy solid-liquid coexistence simulations,[79] and the variation tendency in pressure and temperature is in good agreement with that observed in experiments.[80]

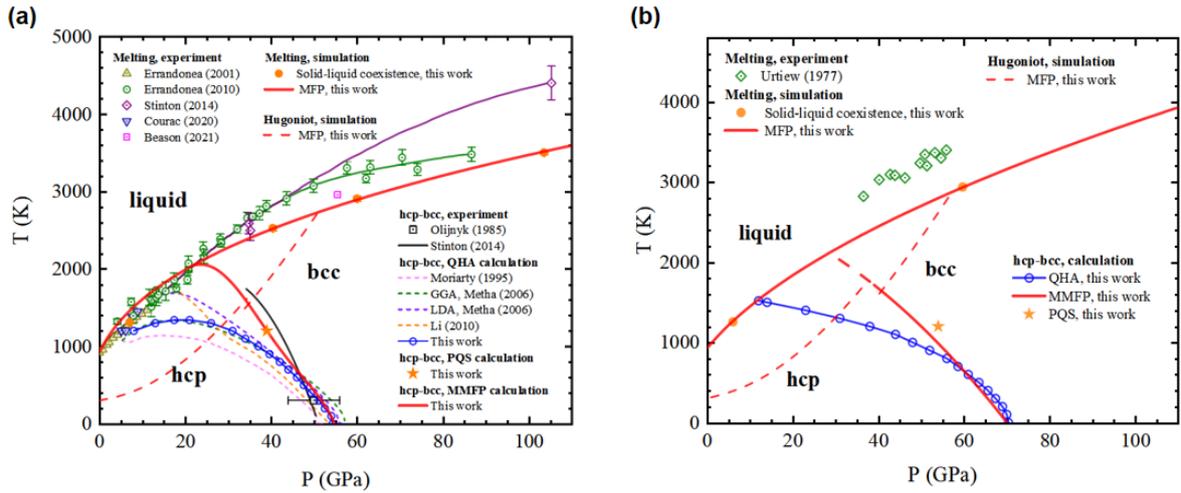

**Figure 5.** Phase diagram of (a) Mg and (b) the Mg-Al alloy AZ31B. Comparison between the results calculated using MFP$^2$ with other first-principles calculations and experimental measurements.

## 3.2. HTEM for Thermo-elastic Properties

HTEM[61–63] is an efficient computational tool for predicting elastic constants, elastic moduli, and sound velocities of materials at high temperatures based on first-principles calculations. It employs a computational framework combining the Quasi-Static Approximation (QSA)[81] with the energy-strain method, and can also determine high-temperature elastic properties using the stress-strain method based on molecular dynamic simulations. Utilizing a proposed semi-analytical model for high-temperature and high-pressure elasticity,[62] HTEM can predict elastic properties over a wide temperature and pressure range using data from only a few selected volumes. HTEM also provides the capability for visualizing the evolution of elasticity and its anisotropy with temperature and pressure, showing intuitively how mechanical properties change under different conditions. HTEM consists of two core functional modules: (1) Isothermal elastic constant calculation module and (2) Isentropic elastic constant conversion module. Users only need to provide standard VASP input files and a state input file containing necessary thermodynamic parameters, and an automated workflow can be conducted to calculate high-temperature elastic properties. Tests on Mg and typical Mg-Al alloys showed that the predicted longitudinal and bulk sound velocities deviate by less than 8% from experimental measurements,[75, 82, 83] confirming the accuracy of HTEM in predicting the elastic properties at high temperatures and high pressures. HTEM has also been successfully applied to study the high-temperature elastic properties of titanium-based high-entropy alloys.[63]

### 3.3. TREX for Electrical and Thermal Conductivity

TREX[64, 65] is a software designed to predict the electrical and thermal conductivity of materials at high temperatures. It can be applied to calculate the electro-thermal transport properties of diverse materials, including metals, alloys, and ceramics, across wide temperature and pressure conditions. TREX consists of two core functional modules: (1) The electronic transport module, which performs high-accuracy calculations of electrical conductivity and electronic thermal conductivity based on the Kubo-Greenwood formula,[84] using atomic configurations generated from molecular dynamics simulations and their corresponding calculated electronic structure as input. (2) The lattice thermal conductivity module, which rapidly evaluates the lattice thermal conductivity using the empirical Slack equation,[85] with basic thermodynamic properties from experiments or first-principles calculations used as input. Simulations of the electrical and thermal conductivity of Mg and typical Mg-Al alloys as a function of temperature up to 50 GPa[65] show maximum relative errors of less than 20% compared with experimental measurements.[65, 86–88] The methodology in TREX has also been effectively used to investigate anomalous variations in the electrical resistivity of high-entropy alloys.[89]

### 4. Hippos for Phase-Field Simulations under High Strain Rates

The phase-field method provides a powerful mesoscale framework for simulating the evolution of material microstructures. Coupling with continuum mechanics constitutive models enables the simulation of complex processes such as twinning and recrystallization, including their nucleation and growth, thereby linking microstructural changes to the material's mechanical response, particularly during dynamic loading.[90–94] However, accurately modeling behavior under high strain rate conditions presents specific challenges. These include determining appropriate kinetic coefficients for twin growth, capturing the coupled evolution of twinning and recrystallization microstructures, and simulating the nucleation and growth of grains from the liquid phase should melting occur. To address these specific requirements, we have developed an advanced phase-field model tailored for microstructure evolution under high strain rate loading[95] and implemented it in the corresponding simulation software, Hippos.

**4.1. Calculation of Twinning Kinetic Coefficients**

The growth rate of twins is influenced by two factors: the externally applied strain rate and the microscopic stress distribution. Experimental results show that twin boundary growth speeds vary over a wide range, from $10^{-9}$ to $10^2$ m/s, as the strain rate increases.[96] Therefore, it is necessary to establish a kinetic coefficient that depends on the system's average stress, such that when the input strain rate parameter changes, both the stress-strain curve and the twin interface evolution speed change accordingly. By matching the relationship between stress and twin growth velocity observed in molecular dynamics simulations (Figure 6(c)), we determined the kinetic coefficient $L$ for the phase-field order parameter evolution equation. Subsequently, using the improved twinning phase-field model, we studied the nucleation and growth process of a single twin variant in a single crystal, outputting the stress-strain curve for comparison with experimental results. The simulations show that at stresses of 200 MPa (Figure 6(d) I) and 350 MPa (Figure 6(d) III), the twin growth speeds are 8 m/s and 12 m/s, respectively, consistent with molecular dynamics results. Twin growth in a single crystal typically proceeds in three stages: First, slow twin growth with continuously increasing stress. Second, the twin growth speed increases with stress, leading to stress release within the grain and a decrease in the stress-strain curve. Third, twin growth reaches its limit, and stress within the grain increases again. Figure 6(e) shows the integrated calculation workflow for Hippos within the ProME platform: Elastic constants at different temperatures are obtained from the HTEM software and input into the Hippos twinning module. Hippos simulates the microstructure evolution, obtaining the spatial distribution of grain structure order parameters and solving for dislocation density. The dislocation density parameters are then interpolated and post-processed to derive stress-strain curve information, which is finally passed to a constitutive model (e.g., JC-P4) for fitting. The phase-field model can be referred to in the Supporting Information.

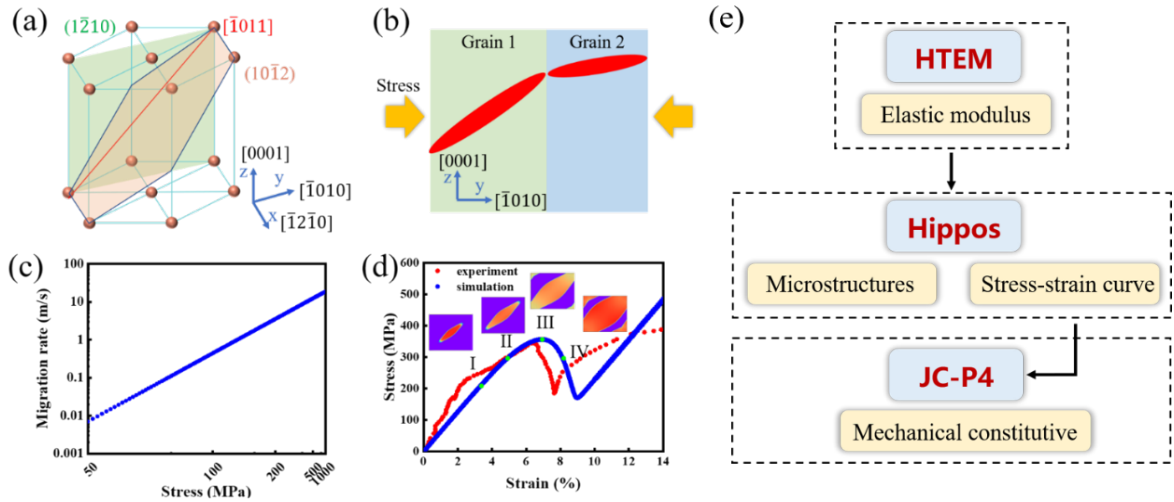

**Figure 6.** Schematic of the $(10\bar{1}2)[\bar{1}011]$ twinning system in adjacent grains and simulated stress-strain curves for twin growth. (a) The $(10\bar{1}2)[\bar{1}011]$ twin system, highlighting the $(1\bar{2}10)$ cross-section (green) used for the simulation. (b) Schematic showing twin orientations in adjacent grains. Arrows indicate the direction of applied compressive stress. (c) Twin boundary growth speed as a function of stress, determined from molecular dynamics (MD) simulations. (d) Simulated stress-strain curve for the growth of a single twin variant within a single crystal. (e) Flowchart illustrating the Hippos integrated calculation workflow.

### 4.2. Phase-Field Modeling of Coupled Twinning and Recrystallization

Studies indicate that during dynamic loading at strain rates in the range of 10 to $10^4$ s$^{-1}$, twinning preferentially occurs during the initial stages of deformation. As deformation increases, a coupled mechanism involving both twinning and recrystallization emerges.[97-99] Twinning dominates initially primarily because the high strain rate induces significant plastic deformation within a very short loading time. Dislocations within the material cannot respond quickly enough, necessitating the twinning mechanism to accommodate the concentrated stress. Twinning often completes its nucleation and growth process during the small strain stage (before reaching the critical strain for recrystallization nucleation). Research suggests that twin boundaries provide numerous nucleation sites for subsequent dynamic recrystallization, a mechanism termed twin-induced dynamic recrystallization (TDRX).[100, 101] TDRX has been widely reported in systems like magnesium and titanium alloys, although the precise physical mechanisms are still under investigation.[102-105]

Figure 7 illustrates the computational workflow for coupled twinning-recrystallization evolution: After inputting the initial crystal structure and relevant parameters, a visco-elasto-plastic model is used to solve for elastic-plastic deformation. When the stress exceeds the critical resolved shear stress for twinning activation, twins begin to nucleate and grow within the polycrystal. Upon reaching a specified deformation level, the twinned microstructure is output and used as input for simulating dynamic recrystallization during subsequent large deformation stages. Twin interfaces provide preferential nucleation sites, facilitating the nucleation and growth of dynamically recrystallized grains. Hippos can simulate the microstructural evolution during TDRX. By combining this with post-processing using plasticity constitutive models like pseudo-slip[106] and Kocks-Mecking models,[107] it can output the corresponding stress-strain curves.

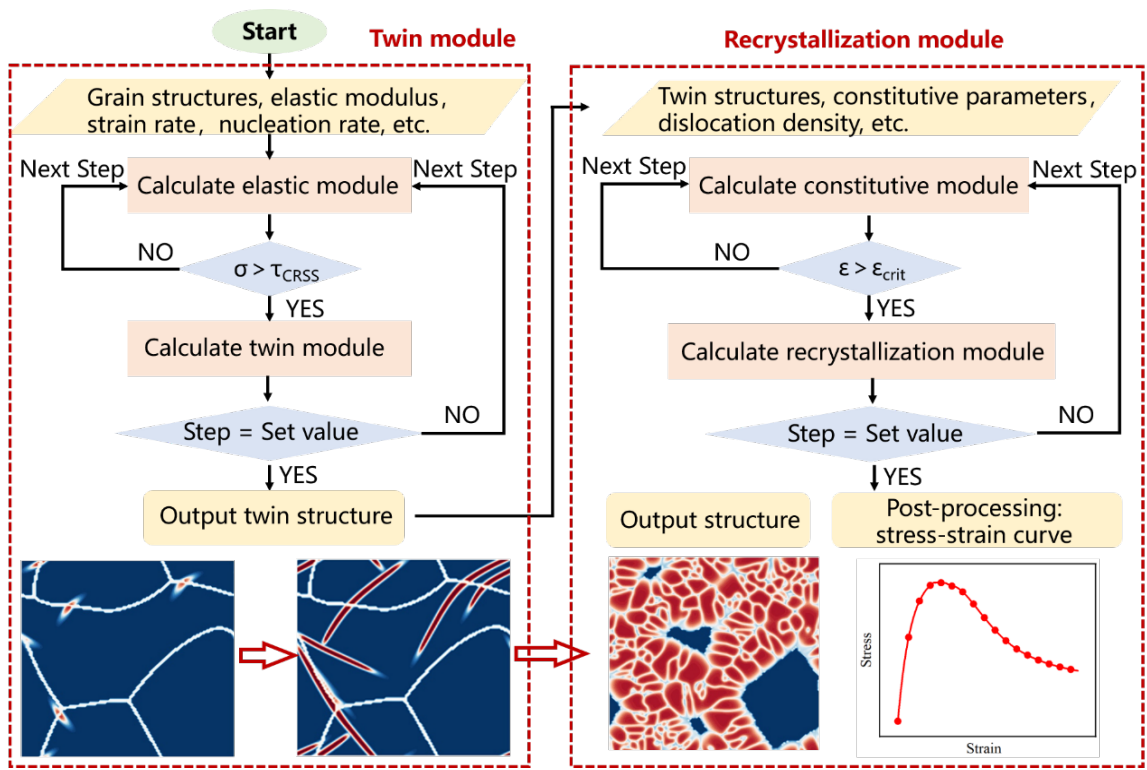

**Figure 7.** Phase-field modeling of coupled twinning and recrystallization.

## 4.3. Phase-Field Model and Validation for Melting and Recrystallization under High Strain Rates

Light-gas gun shock wave experiments can achieve strain rates exceeding $10^6$ s$^{-1}$ and are a primary experimental method for exploring the nonlinear response behavior of materials under extreme strain rates. By introducing a liquid phase order parameter, Hippos constructs a multiphase phase-field model comprising liquid, polycrystalline, and recrystallized phases. This enables simulation of the melting of the initial polycrystalline phase under high strain rates, followed by random nucleation and solidification from within the liquid phase during unloading.

A typical magnesium alloy sample is shown in Figure 8(a), with the initial structure containing some pre-existing recrystallized grains. Hippos can generate microstructures with partial initial recrystallization based on the sample state (Figure 8(d)). When the peak pressure in the sample reaches approximately 16.4 GPa, columnar grain structures are observed (Figure 8(b)). Phase-field simulations suggest that, due to the extremely high strain rate, dislocation motion is hindered, leading to the formation of columnar grains to accommodate plastic deformation (Figure 8(e)). At a pressure of approximately 42.2 GPa, the recovered sample exhibits evidence of melting followed by recrystallization (Figure 8(c)). Under such ultra-high pressure shock loading, the magnesium alloy reaches its melting point. Some grains undergo solid-liquid phase transformation, entering a molten state. Upon shock unloading, random nucleation of new grains occurs within the liquid phase, as shown in Figure 8(f).

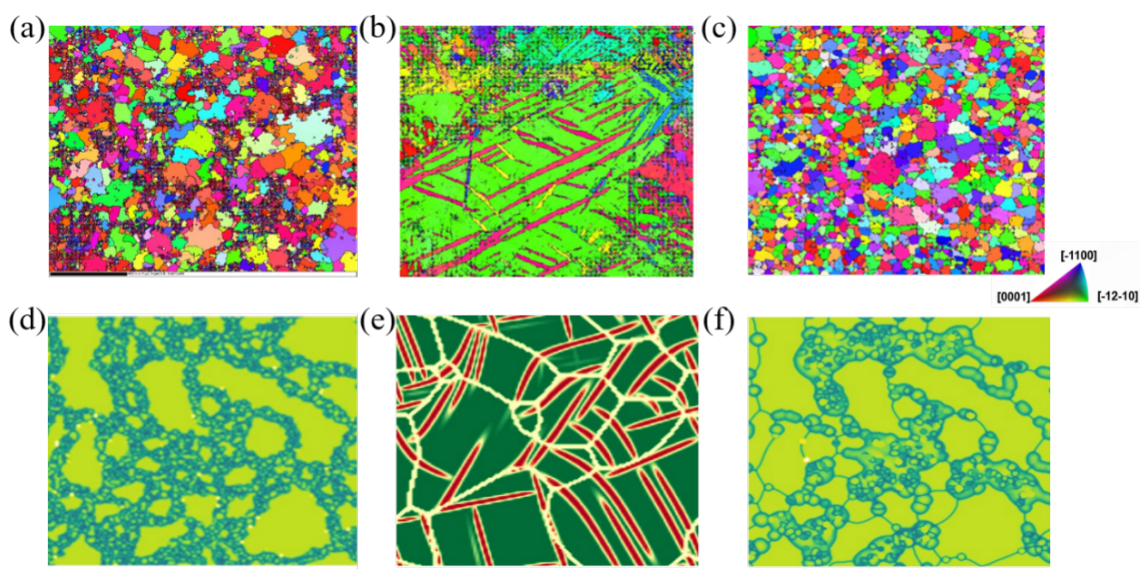

**Figure 8**. Comparison of EBSD characterization from shock-recovered samples and corresponding Hippos phase-field simulation results. EBSD characterization: (a) Initial state (0 GPa shock pressure). (b) After moderate-pressure loading (~16.4 GPa shock pressure). (c) After high-pressure loading (~42.2 GPa shock pressure). Phase-field simulation results: (d) Simulated partially recrystallized polycrystalline structure. (e) Simulated columnar grain structure. (f) Simulated structure resulting from melting and recrystallization under high pressure.

## 5. AutoCalphad for Integrated Modeling and Automatic Optimization of Phase Diagrams

CALPHAD (CALculation of PHAse Diagrams) is a thermodynamic-based method for computing phase diagrams. It represents the thermodynamic properties of each phase using mathematical expressions for the Gibbs free energy. By combining these models with experimental data or theoretical calculation results, CALPHAD enables the prediction of complex phase equilibria and the calculation of thermodynamic properties in multicomponent systems.[108] Traditional CALPHAD approaches face two major challenges: First, thermodynamic modeling primarily relies on discrete experimental thermodynamic data. Data from different sources often exhibit systematic discrepancies, leading to reduced model reliability. Furthermore, experimental measurements under extreme conditions like high temperature and high pressure are rare due to high costs or technical difficulties. Second, existing commercial software (e.g., Thermo-Calc, Pandat) typically uses least-squares methods to optimize model parameters. This approach can be inefficient, prone to getting trapped in local optima, and susceptible to the "snowball effect" due to strong coupling between parameters, where adjusting one parameter destabilizes the entire model.[109–113] To address these issues, we developed the AutoCalphad software for integrated phase diagram modeling and automatic optimization. AutoCalphad facilitates cross-scale modeling integration and automatic optimization of multiple thermodynamic model parameters, moving beyond the traditional reliance on experimental calibration and manual trial-and-error for multicomponent phase diagram modeling.

**5.1. Integrated Modeling and Calculation of Temperature-Composition Phase Diagrams**

Within the CALPHAD framework, we established an integrated method for modeling multicomponent, multiphase thermodynamic systems. This method couples structure search software (ABC), disordered configuration modeling software (SAE), first-principles calculation software (e.g., VASP), and thermodynamic calculation software ($MFP^2$). By combining models like the Redlich-Kister polynomial and sublattice models to construct Gibbs energy expressions for multicomponent phases, this approach reduces reliance on experimental data and enhances the predictive capability for phase diagrams under extreme conditions.

The workflow for integrated multicomponent phase diagram modeling in AutoCalphad is shown in Figure 9. First, crystal structures are identified using existing databases or the ABC software, and quasi-random atomic configurations are generated using SAE. Energy-volume and electronic density of states-volume relationships calculated by first-principles software are input into the $MFP^2$ software. Based on mean-field potential or Debye models, $MFP^2$ calculates thermodynamic properties such as entropy (S), enthalpy (H), heat capacity (Cp), and Gibbs free energy (G). Next, the MFP2JSON module automatically extracts thermodynamic data from $MFP^2$ output files, converts it into a JSON format readable by AutoCalphad, determines initial model parameters, and generates an initial thermodynamic model file (TDB). Finally, using the generated thermodynamic model, global Gibbs free energy minimization calculations are performed to compute phase diagrams at various temperatures and compositions, yielding information such as phase constituents, phase fractions, phase boundaries, enthalpy of mixing, enthalpy of formation, activities, heat capacity, and Gibbs free energy.

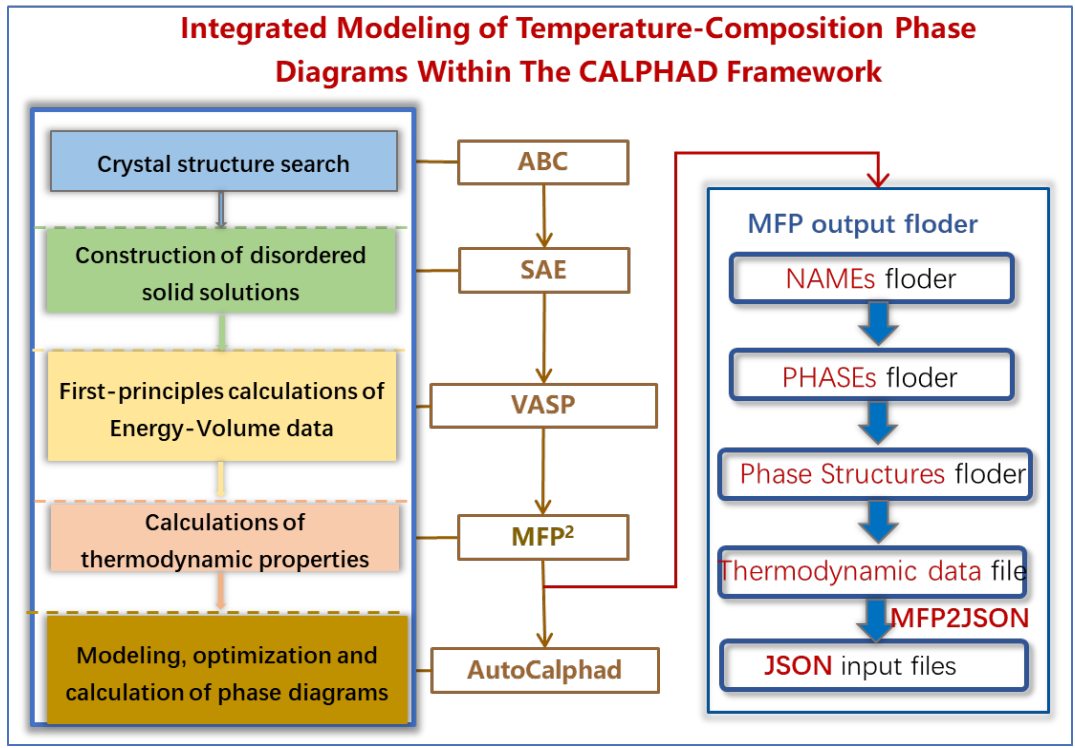

**Figure 9.** The integrated modeling workflow of the temperature-composition phase diagram.

**5.2. Automatic Optimization of Phase Diagram Model Parameters**

Initial CALPHAD models generated solely from first-principles calculation data often exhibit deviations between calculated phase boundaries and experimentally measured ones. This discrepancy arises partly from the inherent errors in density functional theory calculations, which accumulate during parameter transfer and modeling. Therefore, incorporating experimental phase boundary data is necessary to optimize the thermodynamic model parameters. As shown in Figure 10(a), AutoCalphad employs a Markov Chain Monte Carlo (MCMC) algorithm within a Bayesian statistical framework to achieve simultaneous automatic optimization of multiple thermodynamic model parameters while quantifying their uncertainties. Unlike static Monte Carlo random sampling, MCMC dynamically adjusts acceptance probabilities based on Bayesian statistics, reducing inefficient sampling in high-dimensional parameter spaces and identifying high-probability regions of the parameter distribution. This enhances sampling efficiency and significantly reduces the risk of converging to local minima. Furthermore, the MCMC method provides Bayesian-based uncertainty quantification for model parameters and, consequently, for the predicted thermodynamic properties and phase diagrams.

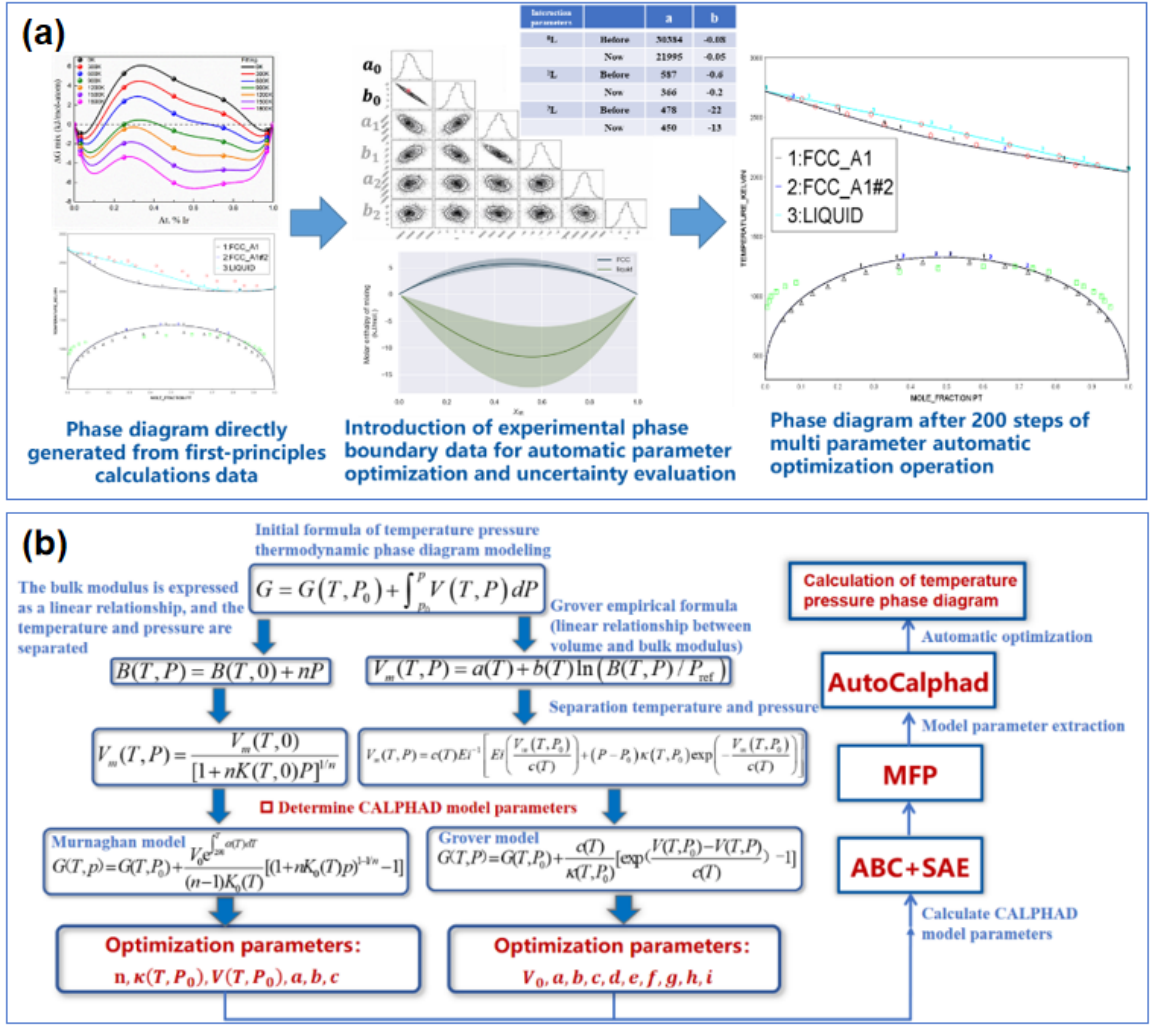

**Figure 10.** (a) CALPHAD model optimization workflow for the Pt-Ir alloy. (b) The integrated modeling workflow of the temperature-pressure phase diagram.

### 5.3. Integrated Modeling and Calculation of Temperature-Pressure Phase Diagrams

Due to the high cost of experimental testing of material properties at high temperatures and pressures, the construction of temperature-pressure (T-P) phase diagrams relies heavily on first-principles calculation data. We developed an integrated modeling method specifically for T-P phase diagrams. This method extracts necessary parameters from high-temperature, high-pressure thermodynamic results obtained via first-principles calculations to establish an initial T-P phase diagram model. This approach partially overcomes the challenge of calibrating high-temperature, high-pressure phase diagrams solely based on experimental data. Subsequent parameter optimization further improves the accuracy of the T-P phase diagram model.

As illustrated in Figure 10(b), the process begins by analyzing Gibbs energy models dependent on temperature and pressure, such as the Murnaghan[113] and Grover[114] models, to identify the required model parameters. Thermodynamic properties of phases at various temperatures and pressures are calculated using the ABC, SAE, and MFP$^2$ software suite. Based on these calculations, parameters like compression coefficients and volume as functions of temperature and pressure are automatically extracted and fitted to generate an initial T-P phase diagram model. Finally, experimental phase boundary data are used to optimize the model parameters, yielding the final calculated T-P phase diagram. Compared to phase diagrams determined directly from first-principles calculations, the deviation of phase boundaries from experimental data can be reduced from around 20% to within 5%.

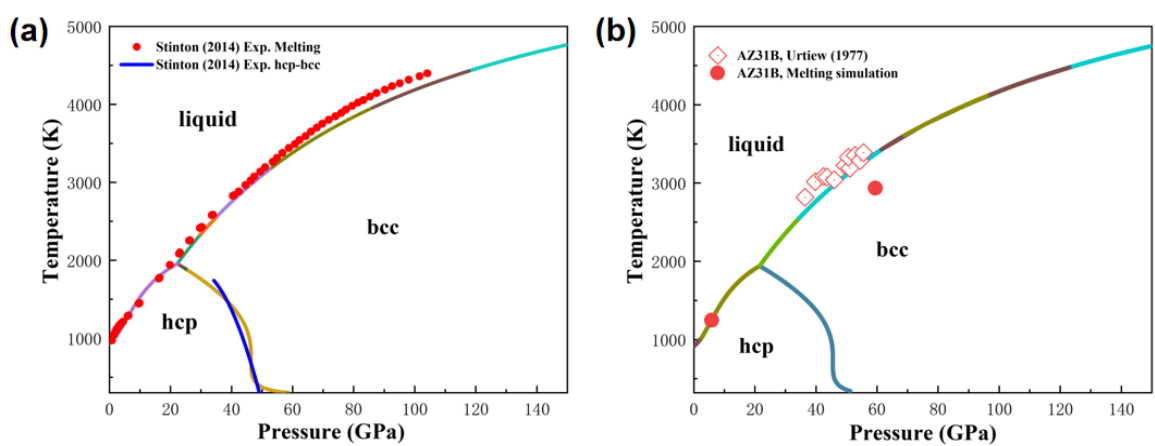

**Figure 11.** Phase diagram of (a) Mg and (b) the Mg-Al alloy AZ31B. Comparison between the results calculated using AutoCalphad with first-principles calculations and experimental measurements.

## 6. The Integrated Computational Platform ProME and Its Application

To effectively integrate the software components described above, we developed the CAST (Computational Alloy Smelting Toolkit) workflow middleware software. CAST decomposes complex workflows into atomic tasks, each with independent objectives and constraints, and maps them onto available computational resources for execution, enabling dynamic scheduling and management.[115] Its built-in fault tolerance module continuously monitors calculation outputs for anomalies and autonomously triggers predefined handling strategies to dynamically adjust the workflow.[116] Building on this infrastructure, CAST supports the creation of master workflow templates using near-natural language semantics, covering the entire process from "structural information -> zero-temperature properties -> temperature-dependent properties -> extreme properties". As shown in Figure 12, a typical workflow starts by using ABC and SAE to predict phase structures and generate atomic configurations from elemental composition information. This input is passed to the CAST middleware, which invokes first-principles software like VASP to perform zero-temperature property calculations over a wide pressure range. The resulting pressure-dependent lattice structures, energies, electronic densities of states, and other ground-state properties are then passed to MFP$^2$ for calculating multiphase thermodynamic properties. These can be further processed by HTEM and TREX to obtain thermoelastic and thermal conductivity properties. The calculated temperature-dependent properties can be compiled into an extreme properties database, passed to AutoCalphad to build thermodynamic models for phase diagrams, or used by Hippos to simulate mechanical response characteristics under shock wave loading (high strain rates) and evaluate impact mechanical performance.

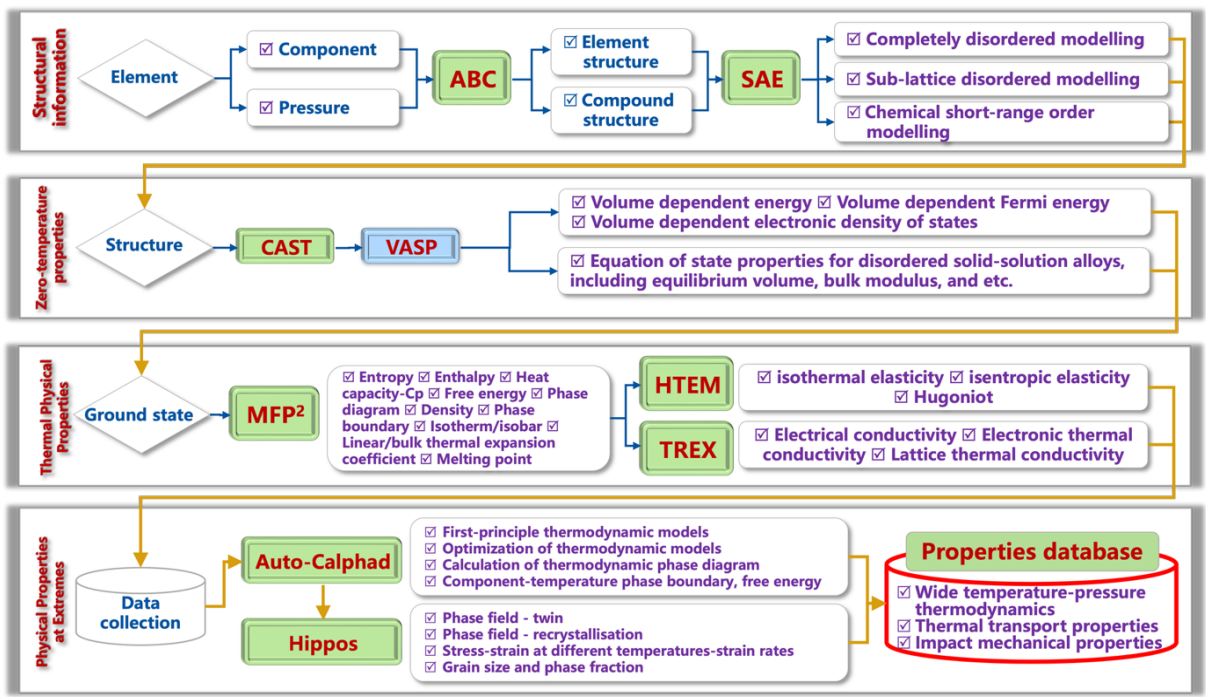

**Figure 12.** The multiscale integrated calculation workflow implemented in ProME, spanning from "structural information -> zero-temperature properties -> temperature-dependent properties -> extreme properties".

The nozzle materials employed in aerospace attitude/orbit control engines are required to exhibit exceptional resistance to high-temperature oxidation at 1,300°C. While conventional binary noble metal alloys have long been the standard for such applications, their prohibitively high material costs present a significant limitation. To address this challenge, this study focuses on the design of a novel quaternary noble metal alloy system, prioritizing both cost efficiency and sustained thermomechanical performance. The proposed compositional optimization strategy aims to achieve "performance - equivalent reduction" in noble metal content, thereby mitigating reliance on expensive raw materials without compromising operational requirements.

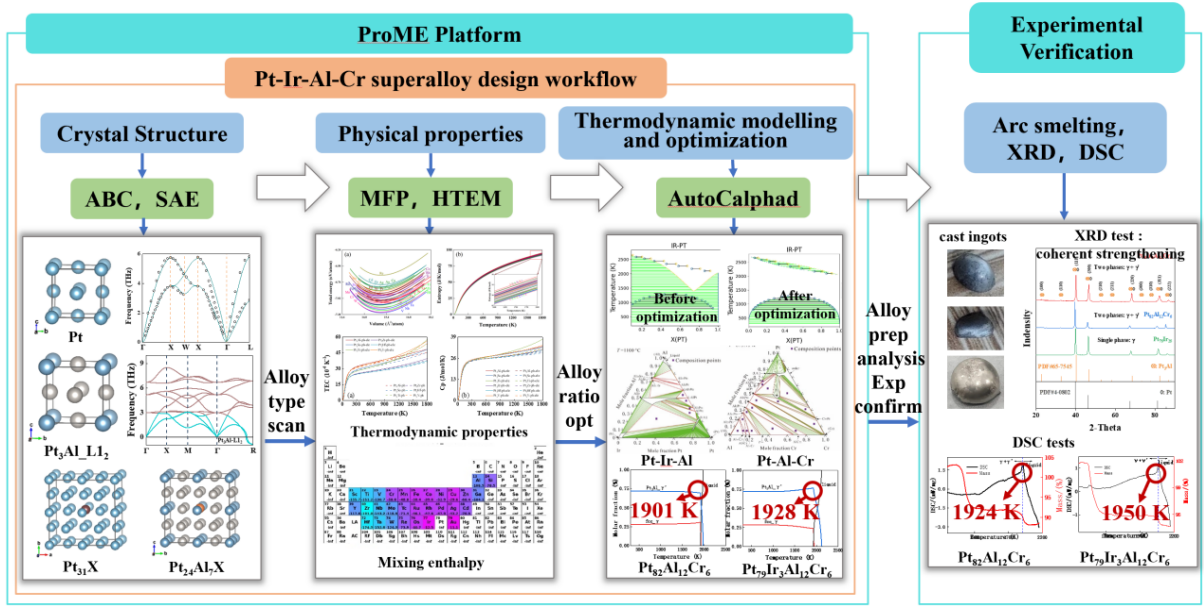

**Figure 13.** ProME application: composition optimization of the Pt-Ir-Al-Cr alloy.

Employing the ProME platform, we completed the compositional design of a Pt-based superalloy for aerospace attitude and orbit control engine nozzles within six months. First, ABC and SAE were used to model solid solutions formed by 33 potential alloying elements with Pt ($\gamma$ phase) and Pt3Al ($\gamma'$ phase). Next, MFP$^2$ and HTEM were used to calculate basic thermodynamic and elastic properties. This screening identified Ir and Cr as elements that effectively strengthen the Pt matrix ($\gamma$ phase) while stabilizing the Pt3Al ($\gamma'$ phase),[117] thus defining the Pt-Ir-Al-Cr quaternary alloy system.[118–123] The thermodynamic properties of all relevant phases within this quaternary system were then calculated. Using these calculated properties, AutoCalphad constructed the thermodynamic model for the Pt-Ir-Al-Cr quaternary alloy. The relationship between Ir and Al content and the $\gamma'$ phase decomposition temperature was calculated, leading to the optimization and identification of the Ir and Al composition range corresponding to a $\gamma'$ decomposition temperature above 1300°C. Based on the computationally designed formulations, Pt-Ir-Al-Cr quaternary alloys were experimentally fabricated. XRD and DSC measurements verified that the γ' phase decomposition temperature closely matched the calculated results. Ultimately, two novel Pt-based superalloys were designed. Their high-temperature strength is comparable to currently used Pt-Ir alloys, but the content of the precious metal Ir was reduced from ~20 at.% to ~3 at.%, potentially lowering raw material costs by approximately 40%. This achieved the design goal of "equivalent reduction" for precious metal alloys (Figure 13).

## 7. Conclusions and Outlook

In this work, we have developed key techniques for simulating material properties under extreme conditions, including high-pressure phase structure modeling, rapid calculation of high-temperature thermodynamic properties, phase-field simulation of microstructure evolution under high strain rates, and thermodynamic modeling of the phase diagrams of multicomponent and multiphase alloys, which lead to the formation a systematic methodology for calculating material properties under extreme conditions. On this basis, we have developed a series of simulation software packages, and have integrated them to build the computational platform ProME (Professional Materials at Extremes) v1.0, which enables integrated calculations of the properties of multicomponent alloys under extreme conditions, covering high temperatures up to tens of thousands of Kelvin, high pressures up to millions of atmospheres, and high strain rates up to millions per second.

For high-pressure phase modeling, we introduced a novel symmetry-constrained configuration sampling algorithm, an on-the-fly MLFF modeling method, and a three-stage relaxation strategy. We developed the ABC software for crystal structure search and combined it with the SAE software for modeling disordered configurations.

For the simulation of high-temperature thermodynamic properties, we developed a new model for rapid free energy calculations based on the MFP method. This led to the MFP$^2$ software, which properly describes both the low-temperature quantum effects and the high-temperature anharmonic effects, with a computational cost comparable to the QHA method, thus achieving good balance between accuracy and efficiency for simulating solid-solid and solid-liquid phase boundaries. We also developed other software for thermal properties, including the HTEM software for thermo-elasticity and the TREX software for electrical and thermal conductivity, establishing systematic capabilities for predicting high-temperature properties of materials.

For simulating microstructure evolution across a broad strain rate spectrum ($10^{-3}$ to $10^6$ s$^{-1}$), the Hippos software was developed. Specifically, for intermediate strain rates, it incorporates a phase-field model describing twin-induced recrystallization evolution. For high strain rates, the accuracy of twinning simulation is enhanced by determining kinetic coefficients informed by molecular dynamics simulations. Crucially, Hippos enables the coupled simulation of twinning and recrystallization phenomena. Validation studies involving typical Mg-Al alloys demonstrated that simulations under high strain rates successfully reproduced experimentally observed features, such as columnar grains and melted/recrystallized structures, under relevant high-pressure conditions (16–42 GPa).

For predicting the multicomponent alloy phase diagrams, the AutoCalphad software was developed based on the CALPHAD framework. It implements an integrated modeling approach that couples inputs from crystal structure searching, disordered configuration modeling, and first-principles thermodynamic calculations. This integration significantly reduces the reliance on often sparse experimental data traditionally required for parameterizing thermodynamic models. Furthermore, AutoCalphad employs Markov Chain Monte Carlo (MCMC) algorithms within a Bayesian framework, enabling the simultaneous, automatic optimization of multiple thermodynamic model parameters. This method offers significantly improved efficiency compared to traditional manual trial-and-error parameter optimization techniques.

For the construction of the integrated computational platform, the CAST workflow middleware was introduced. It facilitates dynamic workflow scheduling and incorporates fault-tolerant computation, particularly for zero-temperature property calculations across wide pressure ranges. It supports the creation of comprehensive computational workflows, spanning from "structural information -> zero-temperature properties -> temperature-dependent properties -> extreme properties". Leveraging the resulting ProME platform, we accelerated the compositional design of novel Pt-Ir-Al-Cr quaternary alloys for aerospace attitude and orbit control engine nozzles, realizing the strategic goal of "equivalent reduction" in precious metal content while maintaining performance.

The establishment of the ProME platform does not only address key technical bottlenecks in simulating material properties under extreme conditions, but also promotes the further integration of computational materials science with practical engineering applications. Targeted at material design under diverse extreme conditions, including high temperatures, high pressures, and high strain rates, ProME provides crucial support for advancing both fundamental scientific understanding and industrial innovation in the research and development of materials.

**Supporting Information**

Supporting Information is appended for reference.

## Acknowledgements

We thank Bangkai Zhu and Ke Xu for the assistance in preparing the manuscript.

## Conflict of Interest

The authors declare no conflict of interests.

## Data Availability Statement

The data that support the findings of this study are available from the corresponding author upon reasonable request.

# Supporting Information

*Xingyu Gao, William Yi Wang, Xin Chen, Xiaoyu Chong, Jiawei Xian, Fuyang Tian, Lifang Wang, Huajie Chen, Yu Liu, Houbing Huang, and HaiFeng Song\**

## 1. ABC: on-the-fly machine learning force-field

This on-the-fly machine learning force-field (MSLFF) is a modified version of TensorMD.[1] The main difference is that this MLFF uses a linear energy model

$$E_a = b_a + G_{abkm}c_{bkm}, \qquad (S1.1)$$

where $E_a$ is the energy of atom $a$, $c_{bkm}$ are linear coefficients, $b_a$ represents the intercept term for atom a. While the relatively small parameter space of linear models means their predictivity might not match that of complex neural network models, they offer distinct advantages. Specifically, the model parameters ($c_{bkm}$ and $b_a$) can be determined rapidly by solving a system of linear equations. Furthermore, due to the limited parameter space, robust solutions can often be obtained with comparatively small training datasets. These characteristics make linear models particularly well-suited for integration into applications like structure searching, where rapid force field updates are beneficial.

Although training the linear model using only energies is straightforward, incorporating atomic forces and structural stress information leads to higher-fidelity solutions. The derivation of terms related to forces and stress within the tensor framework is outlined below

$$P'_{abkd} = 2T_{dm}G'_{abkm}P_{abkd}, \qquad (S1.2)$$

$$Z^{(1)}_{abkmxc} = P'_{abkd}M_{abcd}H'_{abck}R'_{abcx}, \qquad (S1.3)$$

$$Z^{(2)}_{abkmxc} = P'_{abkd}M_{abcd}H_{abck}M'_{abcdx}, \qquad (S1.4)$$

$$G^{(f)}_{baxbkm} = \Sigma_c\left(Z^{(1)}_{abkmxc} + Z^{(2)}_{abkmxc}\right), \qquad (S1.5)$$

$$G^{(h)}_{axybkm} = \Sigma_{ac}\left(Z^{(1)}_{abkmxc} + Z^{(2)}_{abkmxc}\right)R'_{abcx}R_{abc}. \qquad (S1.6)$$

The resulting higher-order tensors, $G^{(f)}_{baxbkm}$ and $G^{(h)}_{axybkm}$, represent the linear coefficients relating the fingerprint components to atomic forces (on atom $b$ in direction $x$ due to neighbor $a$) and structural stress (component $xy$), respectively. The summation in Equation S1.5 is over neighbor index $c$, and in Equation S1.6 is over central atom index $a$ and neighbor index $c$. This allows for the construction of a comprehensive system of linear equations that includes energy, force, and stress data, enabling the determination of the coefficients $c_{bkm}$ and $b_a$.

## 2. ABC: the symmetry-constrained sampling algorithm

The algorithm described below generates symmetrically distinct crystal structures for a given atomic composition (e.g., $Mg_1Al_3$) and a target space group (e.g., *Pm-3m*):

1. **Generate Initial Combinations:** Utilizing depth-first search and combinatorial algorithms,[2] generate all possible combinations of Wyckoff positions such that the sum of the multiplicities for all positions within a single combination equals the total number of atoms dictated by the chemical formula.
2. **Apply Occupancy Constraints:** Filter these initial combinations based on constraints related to the degrees of freedom (DoF). For example, Wyckoff positions with zero DoF might be restricted to appear only once per combination. Limits on the number of times a specific Wyckoff position can be reused within a combination can also be applied.
3. **Enforce Compositional Constraints:** Perform a second screening based on the atomic composition, ensuring that within any valid combination, each specific Wyckoff position is occupied by atoms of only a single element type.
4. **Filter by Combination Length:** Apply a third filter based on the "length" of the Wyckoff combination, defined as the number of distinct Wyckoff positions included in the combination. For high-pressure structure searches, this maximum length is often restricted (e.g., typically set to 1 + number of distinct element types).
5. **Verify Target Symmetry:** Conduct a final screening to ensure that the actual highest symmetry (the determined space group) of the crystal structure generated from a given Wyckoff combination precisely matches the *target* space group. For instance, if targeting space group *I-42d* for a hypothetical $Mg_4$ structure, using only the 4a Wyckoff position results in a crystal structure whose actual highest symmetry is *I41amd*. Therefore, this combination would be discarded if the target was strictly *I-42d*.

After identifying all valid symmetric Wyckoff position combinations (or "occupation states"), a sampling weight *P* is assigned to each combination to guide the initial structure generation process. This weight is calculated using the following formula

$$P = \frac{D}{NL}, \tag{S2.1}$$

where
- *N* is the total number of valid Wyckoff combinations identified for the target space group.
- *L* is the length of the current Wyckoff combination (as defined in step 4).

- *D* is a correction factor related to the total structural degrees of freedom (DoF) of the combination. *D* is set to 2 if the total DoF is 1, and *D* is set to 1 if the total DoF is greater than 1.

The rationale for the *D* factor is that when the initial simulation volume is fixed, a structure with only 1 total structural DoF (e.g., only the lattice parameter *a* in a cubic cell) effectively has zero *internal* structural degrees of freedom that need optimization. A classic example is the body-centered cubic (BCC) structure, described by the 2a Wyckoff position in space group *Im-3m*, which has zero internal DoF. This weighting scheme ensures that configurations possessing higher symmetry and fewer degrees of freedom are assigned a higher initial sampling probability. Table S1 presents examples of typical symmetric Wyckoff occupations, total degrees of freedom, and the resulting initial weights for the MgAl$_3$ and elemental Mg systems.

Table S1 Examples of the Wyckoff sites, DoF and initial weight

| System | Space Group | Wyckoff Sites | L | DoF (Cell) | DoF (Atom) | DoF | Initial Weight |
|---|---|---|---|---|---|---|---|
| MgAl$_3$ | Pm-3m | Mg/1a, Al/3c | 2 | 1 | 0 | 1 | 0.2500 |
| | P6/mmm | Mg/1b, Al/3f | 2 | 2 | 0 | 2 | 0.1000 |
| | | Mg/1b, Al/1a2c | 3 | 2 | 0 | 2 | 0.0667 |
| | | Mg/1a, Al/1b2e | 3 | 2 | 1 | 3 | 0.0333 |
| | Pmmm | Mg/1h, Al/1g2t | 3 | 3 | 1 | 4 | 0.0005 |
| Mg$_8$ | Fd-3m | Mg/8a | 1 | 1 | 0 | 1 | 1.0000 |
| | | Mg/8b | 1 | 1 | 0 | 1 | 1.0000 |
| | Pm-3n | Mg/2a6c | 2 | 1 | 0 | 1 | 0.5000 |
| | Pm-3m | Mg/8g | 1 | 1 | 1 | 2 | 0.3333 |

Once the complete set of valid Wyckoff combinations and their corresponding sampling weights (P) are obtained, they guide the generation of initial crystal structures. Distinct combinations are randomly selected without replacement, according to their weights. After an initial sampling phase where each unique combination has been selected once, combinations corresponding to structures with only one degree of freedom (representing structurally determined configurations at a given volume, like BCC, which are considered "unique" in this context) are typically removed from the pool for subsequent selections. The sampling weights for the remaining combinations are then recalculated using Equation S2.1, and the process of random selection without replacement continues with this updated set and renormalized weights, thereby focusing the search effort on structures with internal geometric flexibility.

## 3. MFP²: characteristic temperature in MFP

An efficient representation of the mean-field potential in the MFP method is[3]

$$g(r,V) = \frac{1}{2}[E_c(R+r) + E_c(R-r) - 2E_c(R)] + \frac{\lambda r}{2R}[E_c(R+r) + E_c(R-r)], \quad (S3.1)$$

where $E_c$ is the ground-state energy when all ions are fixed at their equilibrium lattice sites, $r$ is the distance that a single atom deviates from its equilibrium position, $R$ is the Wigner-Seitz radius of the lattice corresponding to the specific volume $V = 4\pi R^3/3$, and $\lambda$ is an integer parameter corresponding to one of the three commonly used expressions of the Grüneisen parameter in MFP.[4-6] The Taylor expansion of Equation S3.1 yields

$$g(r,V) = k_\lambda(V) r^2 + o(r^4), \quad (S3.2)$$

$$k_\lambda(V) = \frac{1}{R^{2\lambda}} \frac{\partial}{\partial R}\left[R^{2\lambda} \frac{\partial E_c(R)}{\partial R}\right], \quad (S3.3)$$

where $k_\lambda(V)$ is the generalized force constants from MFP. The characteristic temperature $\theta_\lambda(V)$ can be given as

$$\theta_\lambda(V) = \frac{\hbar}{k_B} \sqrt{\frac{k_\lambda(V)}{M}}, \quad (S3.4)$$

where $\hbar \equiv h/2\pi$ is the reduced Planck constant, $k_B$ is the Boltzmann's constant, and $M$ is the average mass per atom.

## 4. Hippos: the phase-field model

The total free energy of the twin system contains the bulk free energy, gradient energy and elasto-plastic energy[7]

$$F = F_{cry} + F_{gra} + F_{mec}, \quad (S4.1)$$

where the bulk free energy constructs a double well model between the matrix and the twin, and between different twin variants, which is expressed by the following formula

$$F_{cry} = \int_V \Delta f_{twin}\left(\sum_{g=1}^{M}\sum_{q=1}^{p}\left(\eta_g^q\right)^2\left(1-\eta_g^q\right)^2 + B_\gamma \sum_{g=1}^{M}\sum_{q=1}^{p}\sum_{u\neq q}^{p}\left(\eta_g^q\right)^2\left(1-\eta_g^q\right)^2\right)dV. \quad (S4.2)$$

The gradient energy mainly covers the anisotropic interface energy, which is expressed by the following formula

$$F_{gra} = \int_V \left(\sum_{g=1}^{M}\sum_{q=1}^{p}\sum_{i=1}^{3}\sum_{j=1}^{3}\kappa_{ij}(g,q)\cdot\left(\nabla_i \eta_g^q\right)\cdot\left(\nabla_j \eta_g^q\right)\right)dV. \quad (S4.3)$$

Elasto-plastic energy includes elastic energy and plastic energy

$$F_{mec} = \int_V \frac{1}{2}C_{ijkl}\varepsilon_{ij}^{ela}\varepsilon_{kl}^{ela}dV. \quad (S4.4)$$

The following equation is the governing equation for iteratively solving the twin order parameter $\eta$

$$\frac{\partial \eta_g^q(x,t)}{\partial t} = -L \frac{\delta F}{\delta \eta_g^q}. \quad (S4.5)$$

## 5. AutoCalphad: the T-C model

The relationship between the molar Gibbs free energy of pure component $i$ in phase $\varphi$ (where $\varphi$ = LIQUID, BCC, FCC) and temperature (T) can be expressed as

$$G_i^\phi(T) = a + bT + cT \ln T + dT^2 + eT^3 + fT^{-1} + gT^7 + hT^{-9}, \quad (S5.1)$$

where a, b, c, d, e, f, g, and h are fitting parameters.

Gibbs Free Energy Expressions for Liquid and Solid Solution Phases

$$G_m = \sum x_i{}^0 G_i + RT \sum x_i \ln x_i + G_m^E. \quad (S5.2)$$

In Equation S5.2, $x_i$ is the mole fraction of element $i$, $G_i$ represents the molar Gibbs free energy of pure element $i$; the $RT\sum x_i \ln x_i$ term represents the ideal entropy of mixing contribution; $G_m^E$ represents the excess Gibbs free energy, which is expressed using the Redlich-Kister polynomial as

$$G_m^E = x_i x_j \sum_{k=0} L^{(k)} (x_i - x_j)^k. \quad (S5.3)$$

$L^{(k)}$ represents the interaction parameter between elements $i$ and $j$

$$L^{(k)} = a + bT + c \ln T + dT^2 + eT^3 + fT^{-1}. \quad (S5.4)$$

The sublattice model is used to describe solid solution phases and stoichiometric phases/compounds. $(A, B)_m(A, B)_n$ represents a phase with two sublattices having a site ratio of m:n, where the first sublattice can be occupied by components A and B.

The general molar Gibbs free energy of phase $\varphi$ ($G_m^\phi$) is described as the sum of three terms: the reference surface energy ($ref G_m^\phi$), the ideal entropy of mixing contribution ($id G_m^\phi$), and the excess Gibbs free energy ($ex G_m^\phi$), represented by Equation S5.6, S5.7, and S5.8, respectively

$$ref G_m^\phi = y_A' y_A'' G_{A:A}^\phi + y_A' y_B'' G_{A:B}^\phi + y_B' y_A'' G_{B:A}^\phi + y_B' y_B'' G_{B:B}^\phi, \quad (S5.5)$$

$$id G_m^\phi = mRT(y_A' \ln y_A' + y_B' \ln y_B') + nRT(y_A'' \ln y_A'' + y_B'' \ln y_B''), \quad (S5.6)$$

$$ex G_m^\phi = y_A' y_B' \left( y_A'' \sum_i {}^i L_{A,B:A}^\phi (y_A' - y_B')^i + y_B'' \sum_i {}^i L_{A,B:B}^\phi (y_A' - y_B')^i \right)$$

$$+ y_A'' y_B'' \left( y_A' \sum_i {}^i L_{A:A,B}^\phi (y_A'' - y_B'')^i + y_B' \sum_i {}^i L_{B:A,B}^\phi (y_A'' - y_B'')^i \right), \quad (S5.7)$$

$$G_m^\phi = {}^{ref}G_m^\phi + {}^{id}G_m^\phi + {}^{ex}G_m^\phi. \qquad (S5.8)$$

$y_i'$ and $y_i''$ represent the site fractions of component i on the first sublattice and component j on the second sublattice, respectively. $G_{I:J}^\phi$ represents the molar Gibbs free energy of the end-member compound formed when the first sublattice is occupied purely by component I and the second purely by component J, within the structure of phase φ. $L_{I:J,L}^\phi$ represents the interaction parameter between components I and K on the first sublattice when the second sublattice is occupied purely by component J.